\documentstyle[aps,pre,epsf,twocolumn,floats,twoside,graphicx]{revtex}
\graphicspath{{Figures/}{.}}
\begin{document}
\bibliographystyle{prsty}
\date{\today}
\twocolumn[\hsize\textwidth\columnwidth\hsize\csname
@twocolumnfalse\endcsname

\title{\bf Propagation Failure in Excitable Media}

\author{A. Hagberg\thanks{\tt aric@lanl.gov }}
\address{Center for Nonlinear Studies and T-7,\\ Theoretical Division,
         Los Alamos National Laboratory, Los Alamos, NM 87545}

\author{E. Meron\thanks{\tt ehud@bgumail.bgu.ac.il}}
\address{The Jacob Blaustein Institute for Desert Research and
         the Physics Department, \\ Ben-Gurion University, 
         Sede Boker Campus 84990, Israel}

\maketitle

\begin{abstract}
We study a mechanism of pulse propagation failure 
in excitable media where stable
traveling pulse solutions appear via a subcritical pitchfork
bifurcation. The bifurcation 
plays a key role in that mechanism.  Small perturbations, externally
applied or from internal instabilities, may cause pulse propagation
failure (wave breakup) provided the system is close 
enough to the bifurcation point.
We derive relations showing how the pitchfork bifurcation is unfolded by  
weak curvature or advective field perturbations 
and use them to demonstrate wave breakup. 
We suggest that the recent observations of wave breakup 
in the Belousov-Zhabotinsky reaction induced either by an electric 
field~\cite{TMPG:94} or a transverse 
instability~\cite{MKK:94} are manifestations of this mechanism.
\end{abstract}

\pacs{PACS number(s): 82.20.Mj, 05.45.+r}

\vskip2pc]
\narrowtext

%
%
\section{Introduction} 
Failure of wave propagation in excitable
media very often leads to the onset of spatio-temporal disorder. In the
context of electrophysiology it may lead to ventricular fibrillation.
Numerous studies have appeared in the
past few years demonstrating conditions and mechanisms for failure of 
propagation in excitable and bistable 
media~\cite{CoWi:91,HoPa:91,Karma:93,Be:93,BFHNEE:94,HaMe:94b,MaPa:97}. 
Failure may occur by external 
perturbations or spontaneously by intrinsic instabilities. Experimental 
examples include wave breakups in the excitable Belousov-Zhabotinsky (BZ) 
reaction induced by an electric field \cite{TMPG:94} and by a transverse 
instability~\cite{MKK:94}.

\begin{figure}
{\center 
\includegraphics[width=3.0in]{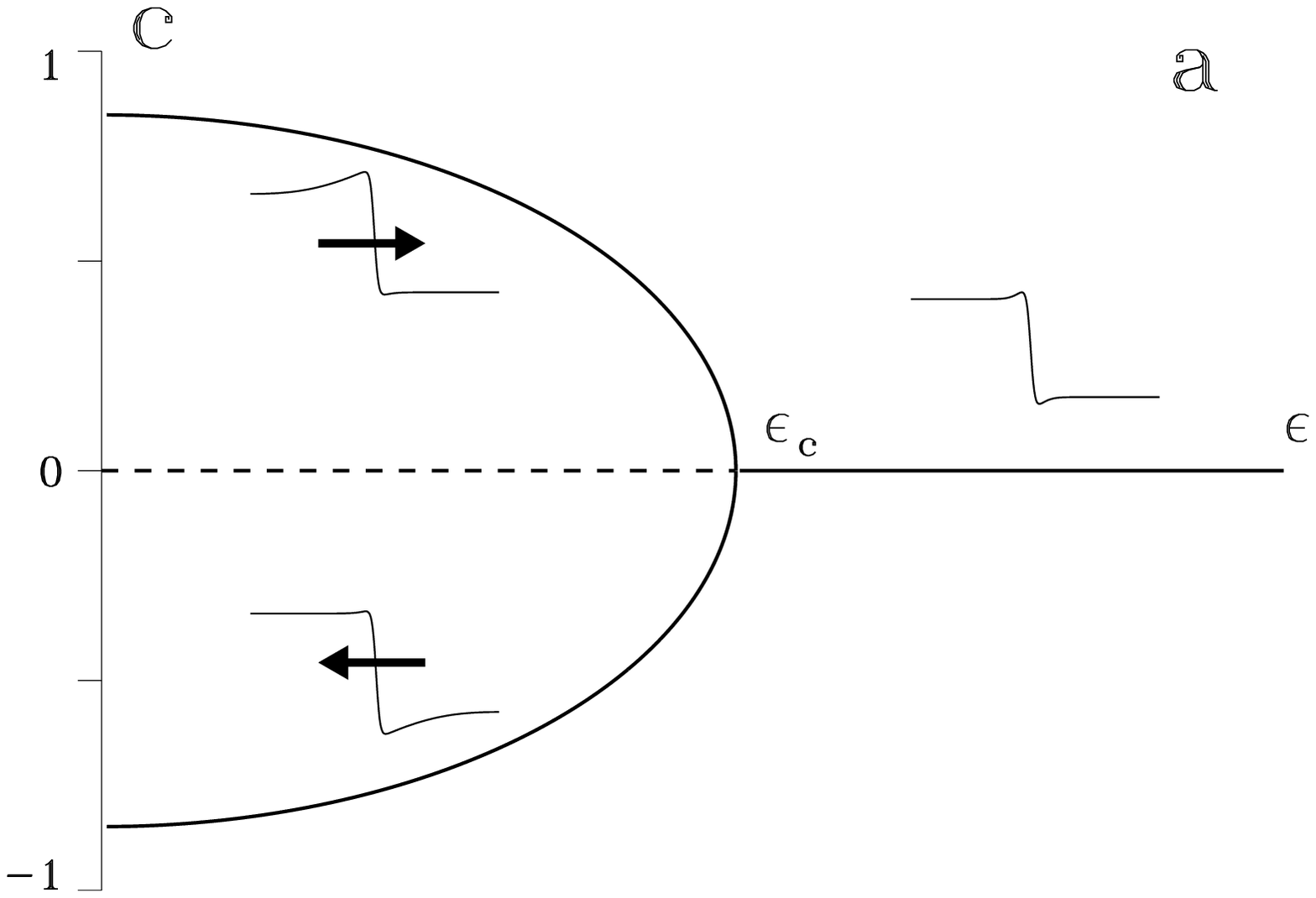}
\includegraphics[width=3.0in]{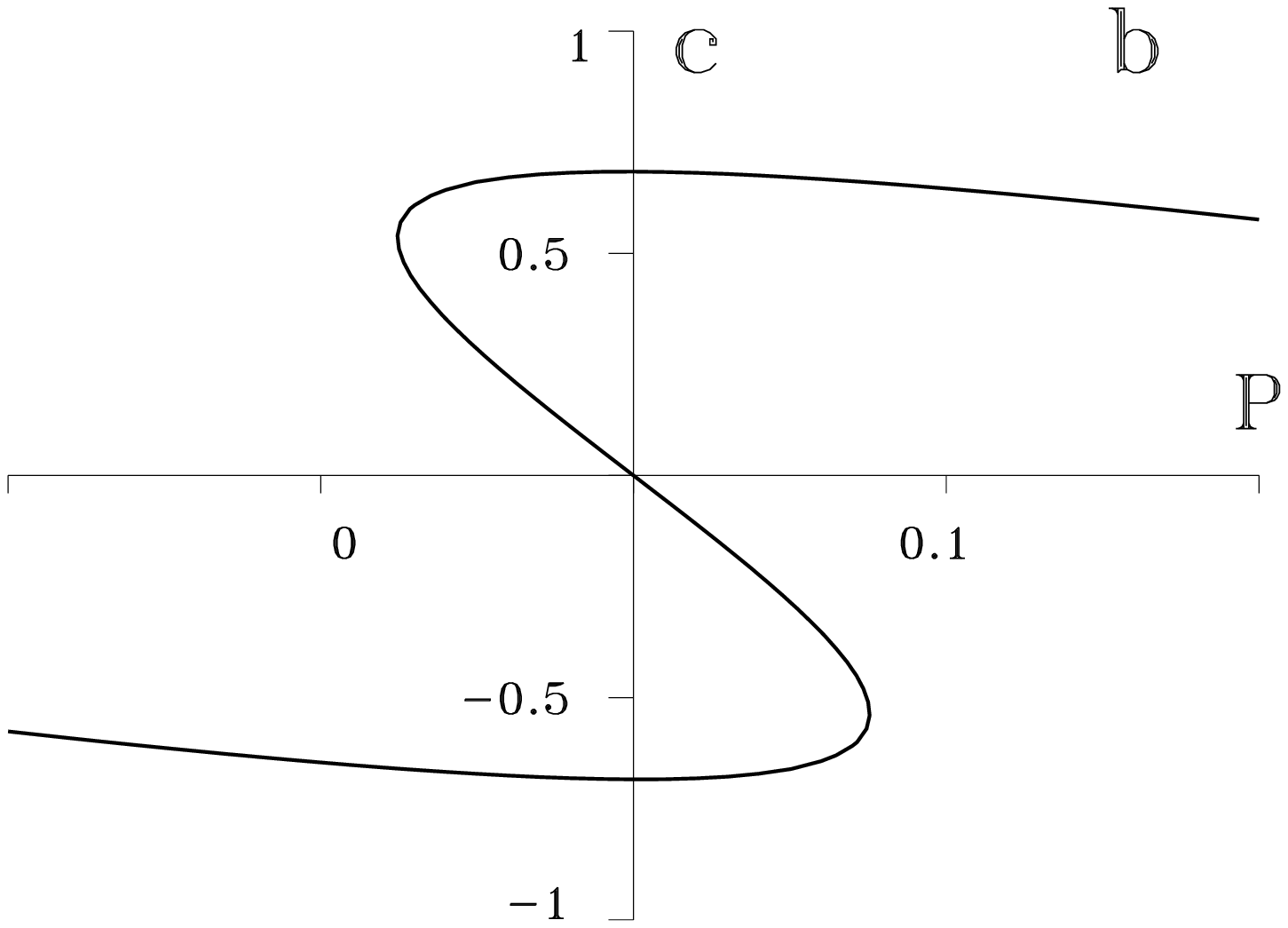} 
}
\caption{Bifurcation diagrams for fronts in the bistable
FHN reaction-diffusion system.
a) At the Nonequilibrium Ising-Bloch bifurcation a stationary
front becomes unstable to a pair of counterpropagating fronts
as a control parameter $\epsilon$ is varied.  The solid
lines represent a branch of front solutions with speed $c$.
b) Unfolding the bifurcation near the critical point, $\epsilon_c$,
gives an S-shaped relation in the unfolding parameter $P$.
}
\label{fig:bifurcation}
\end{figure}

Recently we have attributed domain breakup phenomena in bistable media to 
the proximity to a
pitchfork front bifurcation illustrated schematically in 
Fig.~\ref{fig:bifurcation}a \cite{EHM:95,HaMe:96b}. 
As shown in
Fig.~\ref{fig:bifurcation}b, 
near the bifurcation small perturbations, like an 
advective field or curvature, 
unfold the pitchfork bifurcation to an $S$-shaped relation.
If the perturbation is large enough to drive the system past
the endpoint of a given front solution branch the front reverses
direction.  A local reversal event
along an extended front line in a two-dimensional system
involves the nucleation of a pair of spiral waves and is usually followed 
by domain breakup.  The front bifurcation illustrated in
Fig.~\ref{fig:bifurcation}a 
has been referred to 
in the literature as a Nonequilibrium-Ising-Bloch (NIB) 
bifurcation~\cite{CLHL:90,HaMe:94a}.

\begin{figure}
{\center 
\includegraphics[width=3.0in]{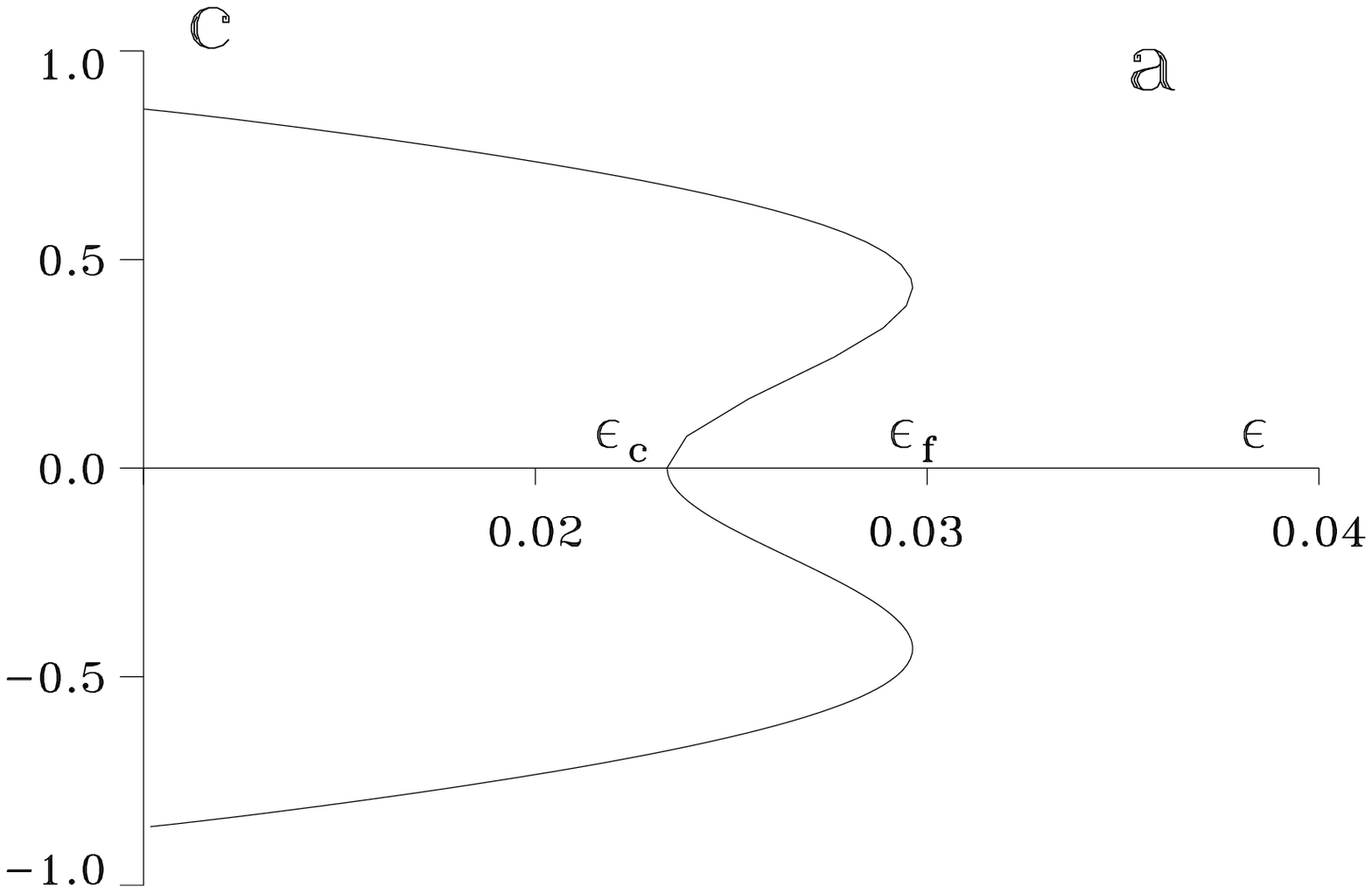}
\includegraphics[width=3.0in]{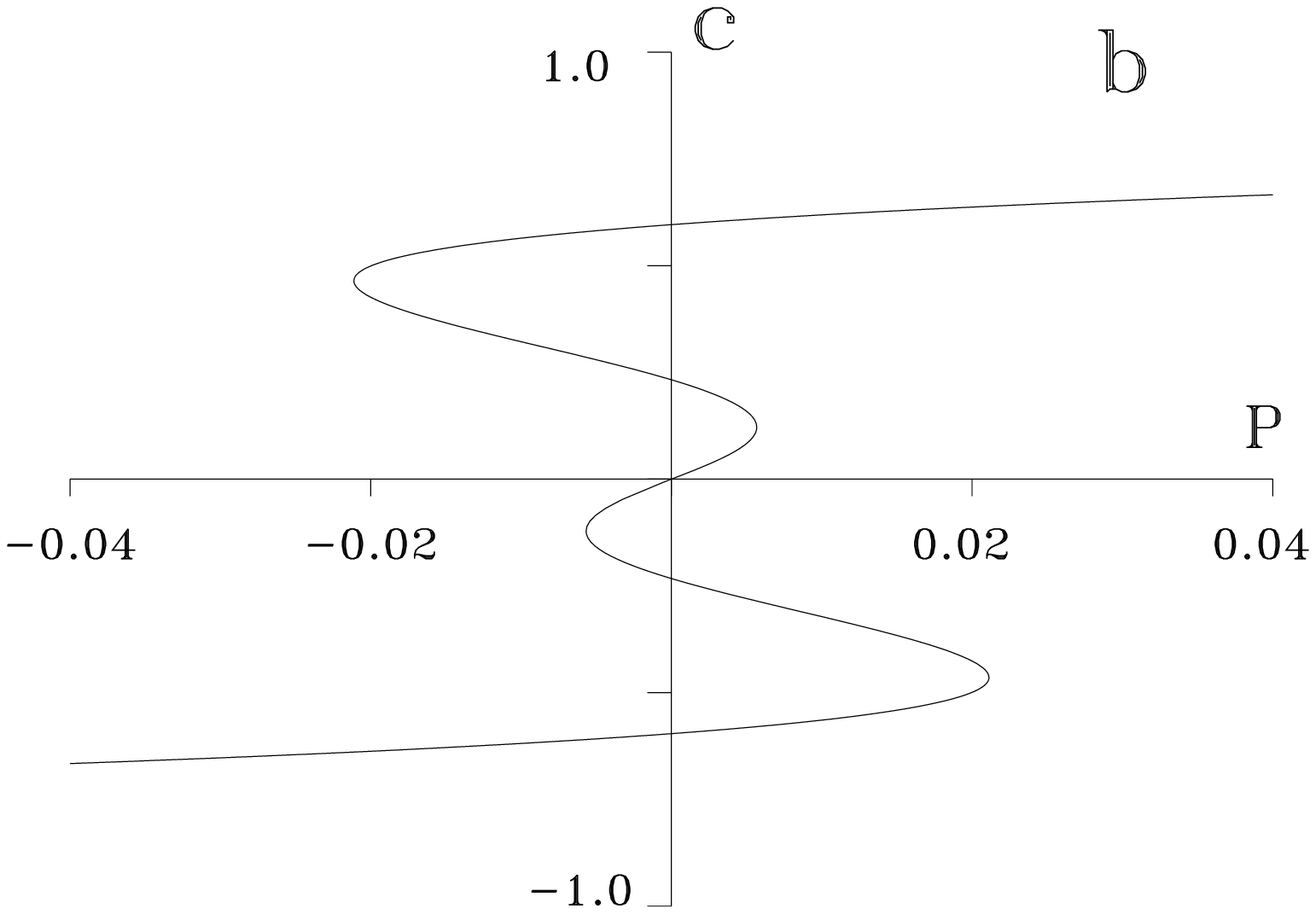}
}
\caption{Bifurcation diagrams for pulse solutions in the 
excitable FHN system.
a) A typical relation for the pulse speed, $c$, vs the 
system parameter $\epsilon$ gives a subcritical pitchfork bifurcation.
b) Unfolding the bifurcation near the critical point, $\epsilon_c$,
gives a multiple S-shaped curve for $c$ in the unfolding
parameter $P$.  }
\label{fig:subcritical}
\end{figure}

In this paper we extend these ideas to wave breakups in excitable media.
The NIB bifurcation is replaced in this case by a subcritical pitchfork 
pulse bifurcation as shown in Fig.~\ref{fig:subcritical}a. 
The typical unfolding of that bifurcation is
shown in Fig.~\ref{fig:subcritical}b. 
Similar to the case of front solutions in bistable systems, 
perturbations that drive the system
beyond the edge point of a given pulse branch may either reverse the 
direction of
pulse propagation, or lead to pulse collapse and convergence to the
stable uniform quiescent state (not shown in Fig.~\ref{fig:subcritical}b). 
The convergence to a uniform attractor is more likely to occur for pulse
structures than for fronts, and has always been observed in our
simulations.
Thus, in our study, the
critical value of the bifurcation parameter, $\epsilon_f$, at which the
upper and lower branches in Fig.~\ref{fig:subcritical}a
terminate, designates failure of 
propagation. Reversals in the direction of propagation (rather than 
collapse) have been observed in experiments on the
Belousov-Zhabotinsky reaction subjected to an electric field~\cite{SMM:94}. 

In two space dimensions a local collapse of a spatially extended
pulse amounts to a wave breakup.
By numerically integrating a FitzHugh-Nagumo (FHN) type model, we
demonstrate two scenarios of wave breakups: breakup induced by an
advective field, modeling an electric field in the BZ reaction, and 
breakup induced by a transverse (or lateral) instability. The two
scenarios, observed both in experiments~\cite{TMPG:94,MKK:94}
and numerical simulations~\cite{TMPG:94,MaPa:97}, reflect the {\em 
same} mechanism: the ability of weak perturbations to
drive transitions from one of the pulse branches to the uniform attractor
when the system is close to failure of propagation. 

In Section II we describe the derivation of a pulse bifurcation diagram
for an FHN model. The derivation applies to both excitable and bistable 
media. The information contained in this diagram is used to draw the 
propagation failure line in an appropriate parameter space. In Section III
we consider the unfolding of the pulse bifurcation by an advective field
and demonstrate wave breakup near the propagation
failure. The unfolding by curvature is studied in Section IV and wave breakup 
induced by a transverse instability is demonstrated. 
We conclude in Section V with a discussion.

%
%
\section{A bifurcation diagram for pulse solutions}

We derive the pulse bifurcation diagram using an activator-inhibitor model
of the FHN type. We assume that the activator varies on a time scale much
shorter than that of the inhibitor and use singular perturbation 
theory~\cite{TyKe:88,DK:89,Meron:92}.
Specifically, we study the pair of equations
\begin{eqnarray}
\label{pde1}
u_t&=&\epsilon^{-1}(u-u^3-v)+\delta^{-1}\nabla^2 u\,, \\
v_t&=&u-a_1v-a_0+ \nabla^2 v \,, \nonumber
\end{eqnarray}
where $\mu=\epsilon/\delta\ll 1$ and the subscripts $t$ denote 
partial derivatives with respect to time. We consider a periodic
wavetrain of planar (uniform along one dimension) 
 pulses traveling at constant speed $c$ in 
the $x$ direction. Each of the excited 
(or ``up state'') domains occupy a length of $\lambda_+$. The recovery 
(or ``down state'') domains are of length $\lambda_-$. In regions where $u$ 
varies on a scale of order unity Eqns.~(\ref{pde1}) reduce to
\begin{eqnarray}
v_{\chi\chi}+cv_\chi+u_+(v)-a_1v-a_0&=0\,,&\qquad -\lambda_-<\chi< 0\,, \nonumber\\   
v_{\chi\chi}+cv_\chi+u_-(v)-a_1v-a_0&=0\,,&\qquad  0<\chi<\lambda_+\,,
\label{outer}    
\end{eqnarray}
where $\chi=x-ct$ and $u_\pm(v)$ are the outer solution branches of the cubic 
equation
$u-u^3-v=0$. For $a_1$ sufficiently large we may linearize the branches 
$u_\pm(v)$ around $v=0$
\begin{equation}
u_\pm(v)\approx \pm 1-v/2\,.\label{linap}
\end{equation}
We solve Eqns.~(\ref{outer}) using the boundary conditions
\begin{eqnarray}
\label{bc}
v(-\lambda_-)&=&v_b\,,\\
v(0)&=&v_f\,,\nonumber\\
v(\lambda_+)&=&v_b\,,\nonumber
\end{eqnarray}
where $v_f$ and $v_b$ are yet undetermined, and the linear approximation 
(\ref{linap}). The solutions are
\begin{eqnarray*}
v&=&A_+\exp(\sigma_1\chi)+B_+\exp(\sigma_2\chi)+v_+\,,\quad -\lambda_-<\chi<
0\,, \\   
v&=&A_-\exp(\sigma_1\chi)+B_-\exp(\sigma_2\chi)+v_-\,,\quad 0<\chi<\lambda_+
\,, \\ 
\end{eqnarray*}
with
\begin{eqnarray}
\sigma_{1,2}&=&-\frac{c}{2}\pm\sqrt{\frac{c^2}{4}+a_1+1/2}\,, \nonumber \\
A_\pm&=&\frac{(v_b-v_\pm)-(v_f-v_\pm)\exp(\mp\sigma_2\lambda_\mp)}
{\exp(\mp\sigma_1\lambda_\mp)-\exp(\mp\sigma_2\lambda_\mp)}\,,\nonumber\\
B_\pm&=&\frac{-(v_b-v_\pm)+(v_f-v_\pm)\exp(\mp\sigma_1\lambda_\mp)}
{\exp(\mp\sigma_1\lambda_\mp)-\exp(\mp\sigma_2\lambda_\mp)}\,,\nonumber
\end{eqnarray}
and $v_\pm=(\pm 1-a_0)/(a_1+1/2)$.
Matching the derivatives of the solutions at $\chi=0$, 
$v^\prime(0^-)=v^\prime(0^+)$, and imposing periodicity on the derivatives, 
$v^\prime(-\lambda_-)=v^\prime(\lambda_+)$, we obtain two 
conditions
\begin{mathletters}
\begin{eqnarray}
\sigma_1 A_+ + \sigma_2 B_+ = \sigma_1 A_- + \sigma_2 B_- \,,
&&\label{cond1}\\
\sigma_1 A_+\exp(-\sigma_1\lambda_-) + \sigma_2 B_+\exp(-\sigma_2\lambda_-) 
&=&\nonumber\\ 
\sigma_1 A_-\exp(\sigma_1\lambda_+) + \sigma_2 B_-\exp(\sigma_2\lambda_+) 
\,.\label{cond2}
\end{eqnarray}
\label{conditions}
\end{mathletters}
Two more conditions are obtained by studying the ``front'' and the ``back'', 
that is,
the leading and trailing border regions between the excited and recovery 
domains. Stretching the spatial coordinate according to $\zeta=\chi/\sqrt{\mu}$
gives the nonlinear eigenvalue problems
\begin{equation}
u_{\zeta\zeta}+c\eta u_\zeta +u -u^3-v_f=0\,,
\label{inner1}
\end{equation}
\begin{eqnarray*}
u(\zeta)\to u_-(v_f)\quad &{\rm as}& \quad \zeta\to\infty\,,\\
u(\zeta)\to u_+(v_f)\quad &{\rm as}& \quad \zeta\to -\infty\,,\\
\end{eqnarray*}
for the narrow front region, and
\begin{equation}
u_{\zeta\zeta}+c\eta u_\zeta +u -u^3-v_b=0\,,
\label{inner2}
\end{equation}
\begin{eqnarray*}
u(\zeta)\to u_+(v_b)\quad &{\rm as}& \quad \zeta\to\infty\,,\\
u(\zeta)\to u_-(v_b)\quad &{\rm as}& \quad \zeta\to -\infty\,,\\
\end{eqnarray*}
for the narrow back region. Here, $\eta=\sqrt{\epsilon\delta}$.
Solutions of (\ref{inner1}) and (\ref{inner2}) yield
\begin{eqnarray}
c\eta&=&-\frac{3}{\sqrt{2}}v_f\,,\label{cond3}\\
c\eta&=&\frac{3}{\sqrt{2}}v_b\,,\label{cond4}
\end{eqnarray}
respectively.

Substituting the relations (\ref{cond3}) and (\ref{cond4})
into Eqns.~(\ref{conditions}) leaves
two equations for the three unknowns $\lambda_+$, $\lambda_-$,
and $c$. The one parameter family of solutions describe periodic 
wavetrains of pulses traveling with speed 
\begin{equation}
\label{c}
c={\cal C}(\lambda;\eta,a_0,a_1)\,,
\end{equation} 
where 
$\lambda=\lambda_+ +\lambda_-$ is the varying wavelength of the family.
The graph of $c$ versus $\lambda$ provides the dispersion relation curve 
obtained in earlier studies~\cite{TyKe:88,DK:89}. 
Our interest here is with the 
behavior of a single pulse and therefore only the limit of large $\lambda$
will be considered. We remind the reader that in deriving these equations 
we have assumed $\mu=\epsilon/\delta\ll 1$ which excludes very small 
$\delta$ values.

\begin{figure}
{\center 
\includegraphics[width=3.5in]{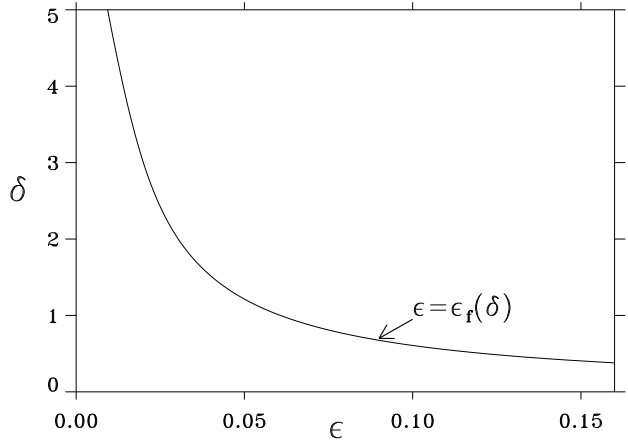}
}
\caption{The pulse failure propagation boundary in the
$\epsilon-\delta$ parameter plane for the excitable FHN model.  
To the right of the line pulses fail to propagate.  
Parameters: $a_1=1.25$, $a_0=-0.2$.
}
\label{fig:epsilon-delta}
\end{figure}

We solved Eqns.~(\ref{conditions}) numerically for both excitable and 
bistable systems at large values of the period $\lambda$.
The solutions were computed by numerical continuation
of known solutions when $a_0=0$, and $\lambda_+=\lambda_-$. 
They yield the typical bifurcation diagram for the speed $c$ 
in terms of the parameter $\epsilon$ as shown in 
Fig.~\ref{fig:subcritical}a.   At some critical value, $\epsilon_f$,
the branch of solutions terminates and past that point
pulses fail to propagate.  The value of $\epsilon_f$
depends on the other system parameters as well.
Figure~\ref{fig:epsilon-delta} 
shows a graph of $\epsilon=\epsilon_f(\delta)$ for an excitable system. 

Note the inherent subcritical nature of the bifurcation \cite{SBG:95,KO:95}.
The bifurcation becomes supercritical only in the 
limit $a_0\to 0$ (pertaining to a bistable medium) where the pulse size tends 
to infinity. In that limit Eqns.~(\ref{conditions})
can easily be solved. The solutions are $c=0$ and $c=\pm 
\frac{2q}{\eta}\sqrt{\eta_c^2-\eta^2}$ for $\eta<\eta_c$, and coincide with 
the Nonequilibrium Ising-Bloch (NIB) bifurcation for front 
solutions~\cite{HaMe:94c}.  

%
%
\section{Wave breakup by an advective field}

\begin{figure}
{\center 
\includegraphics[width=3.0in]{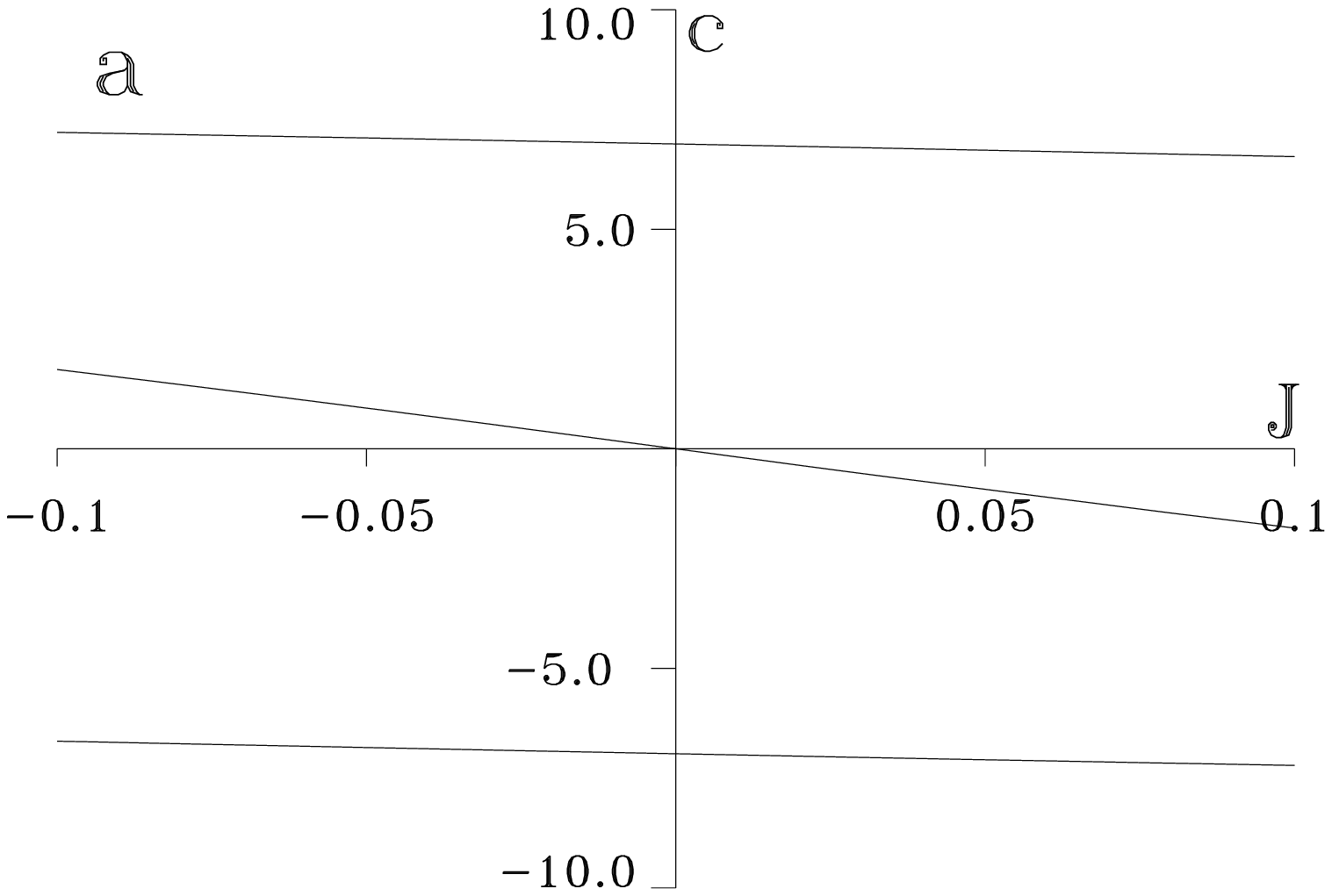}
\includegraphics[width=3.0in]{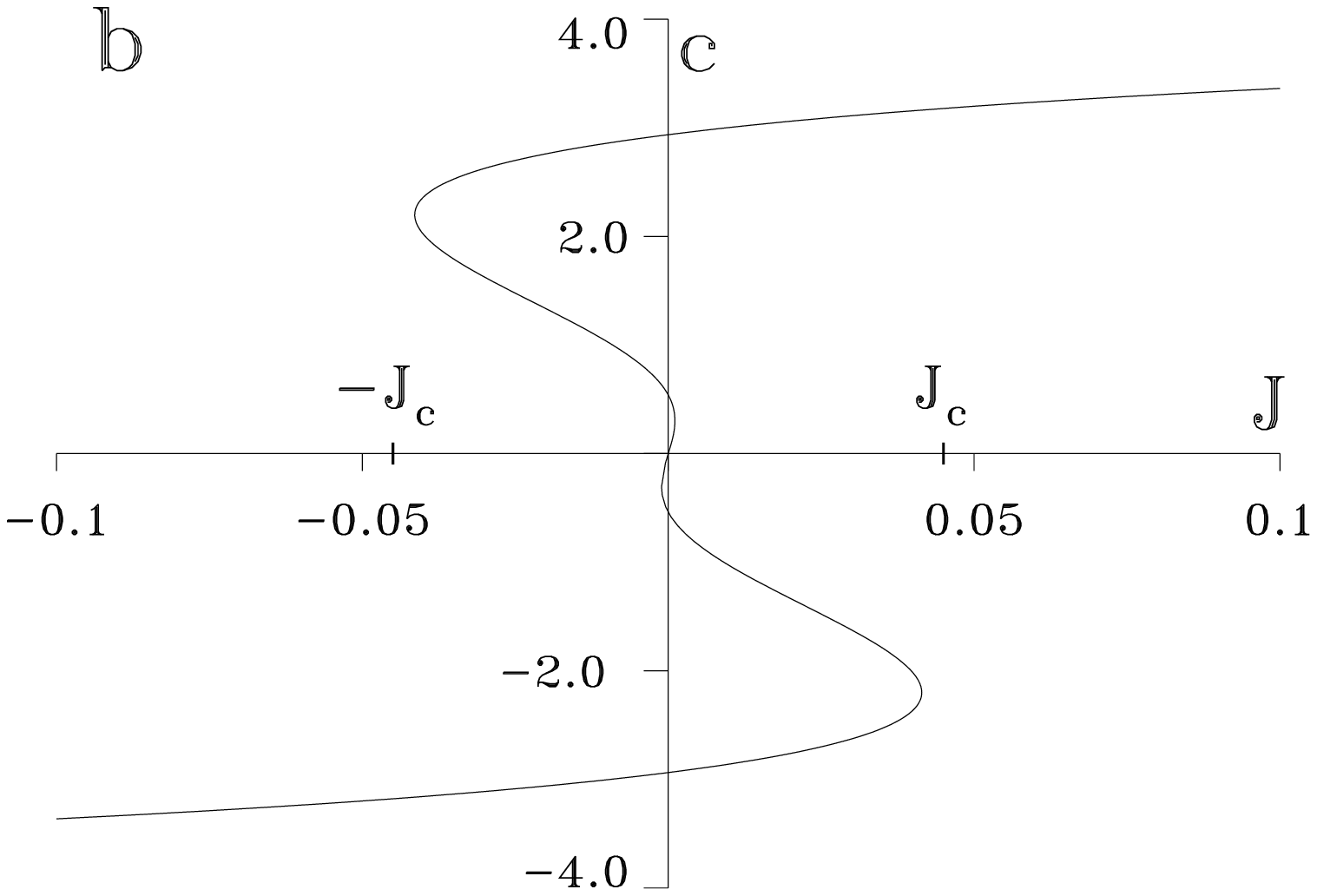}
}
\caption
{
Solutions of the speed, $c$,  vs advection, $J$, relation~(\protect\ref{cJ}).
(a) Away from the bifurcation point, variations in $J$ have little
effect on the pulse speed ($\epsilon=0.01$).
(b) Near the bifurcation point, small variations in $J$ may
drive the system past the endpoint of the solution branch and
cause pulse collapse ($\epsilon=0.05$).
Other parameters: $a_1=1.25$, $a_0=-0.2$, $\delta=1.0$.
}
\label{fig:cvj}
\end{figure}

Application of an electric field to a chemical reaction involving 
molecular and ionic 
species, like the BZ reaction, results in a differential 
advection~\cite{DLK:94}. 
Differential advection in
the FHN equations can be modeled (without loss of generality) by adding an 
advective term to the inhibitor equation
\begin{eqnarray}
u_t&=&\epsilon^{-1}(u-u^3-v)+\delta^{-1}\nabla^2 u\,, \label{pde2} \\
v_t&=&u-a_1v-a_0+ {\bf J}\cdot\nabla v+\nabla^2 v \,, \nonumber 
\end{eqnarray}
where ${\bf J}$ is a constant vector.
Looking for planar solutions propagating at constant speeds in 
the ${\bf J}$ direction, and rescaling 
$\epsilon$ and $\delta$ by the factor $\frac{c}{c+J}$ we find
the influence on the pulse speed by the advection 
\begin{equation}
\label{cJ}
c+J={\cal C}(\lambda;\frac{c}{c+J}\eta,a_0,a_1)\,,
\end{equation}
where ${\cal C}$ is defined in Eqn.~(\ref{c}).
Figures~\ref{fig:cvj}
show the dependence of the pulse speed $c$ on the advection 
constant $J$ obtained by numerically solving Eqn.~(\ref{cJ}) for 
large $\lambda$.  Away from failure of propagation ($\epsilon$ is
significantly smaller than $\epsilon_f$) small variations of $J$ have
little effect on the pulse motion (Fig.~\ref{fig:cvj}a). However, close to failure,
such variations can induce wave breakup by driving the system past the end 
point $J=J_c$  (Fig.~\ref{fig:cvj}b).

\begin{figure}
{\center 
\includegraphics[width=3.25in]{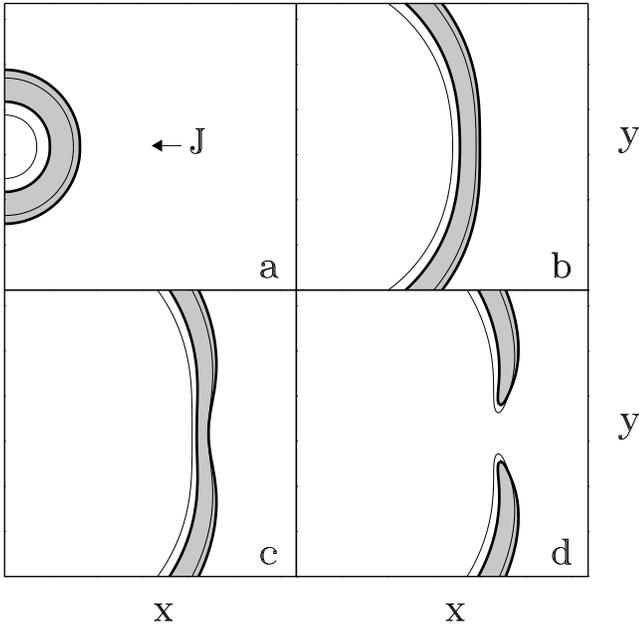}
}
\caption{Numerical solution of Eqns.~(\protect\ref{pde2}) with
a weak advective field ${\bf J}=J{\bf \hat x}$.  The thick and thin 
lines pertain to $u=0$ and $v=0$ contour lines, respectively. The initial
circular pulse fails to propagate along the direction of
the advective field and the pulse breaks.  The pulse continues
to propagate in directions different from the advective field.
The equation parameters are the same as in Fig.~\protect\ref{fig:cvj}.}
\label{fig:jsim}
\end{figure}

Figure~\ref{fig:jsim} shows a numerical simulation 
of Eqns.~(\ref{pde2})
with ${\bf J}=J{\bf \hat x}$ and an initial condition of a curved 
pulse. Along the pulse the effective advection field is the projection of 
${\bf J}$ onto the direction of propagation at that point. 
With $J=0$ the curved pulse propagates uniformly outward in a circular
ring.  Choosing $J$
slightly greater than $J_c$, a wave breakup results: the part of the 
pulse propagating in the ${\bf \hat x}$ direction fails to 
propagate. Those parts propagating in significantly different 
directions still propagate. These results explain earlier 
observations of wave breakup induced by electric 
fields~\cite{TMPG:94}.

%
%
\section{Wave breakup induced by a transverse instability}

Spatially extended pulses, like stripes or disks, may be unstable to transverse
perturbations along the pulse line.  Ohta, Mimura, and Kobayashi 
studied the case of deformations of planar and disk-shaped 
stationary patterns in a piecewise linear FitzHugh-Nagumo
model~\cite{OMK:89}.  Kessler and Levine derived conditions
for the transverse instability of traveling stripes
in a piecewise linear version of the Oregonator~\cite{KeLe:90}.
The curvature induced by a transverse instability can
lead to the formation of labyrinthine patterns~\cite{HaMe:94b,GMP:96,MuOs:96}
or cause spontaneous breakup of a pulse as we will now show.

\begin{figure}
{\center 
\includegraphics[width=3.0in]{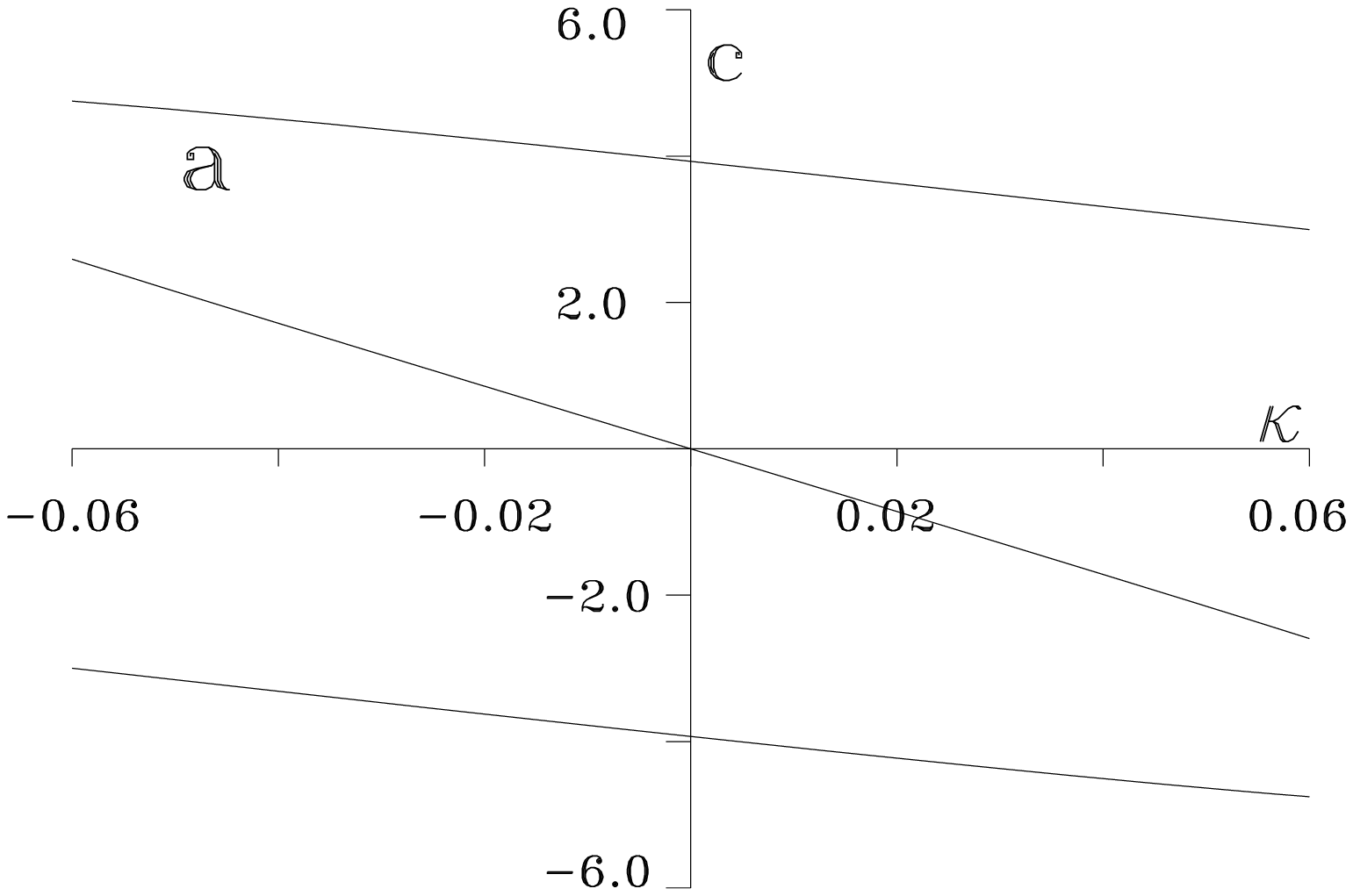}
\includegraphics[width=3.0in]{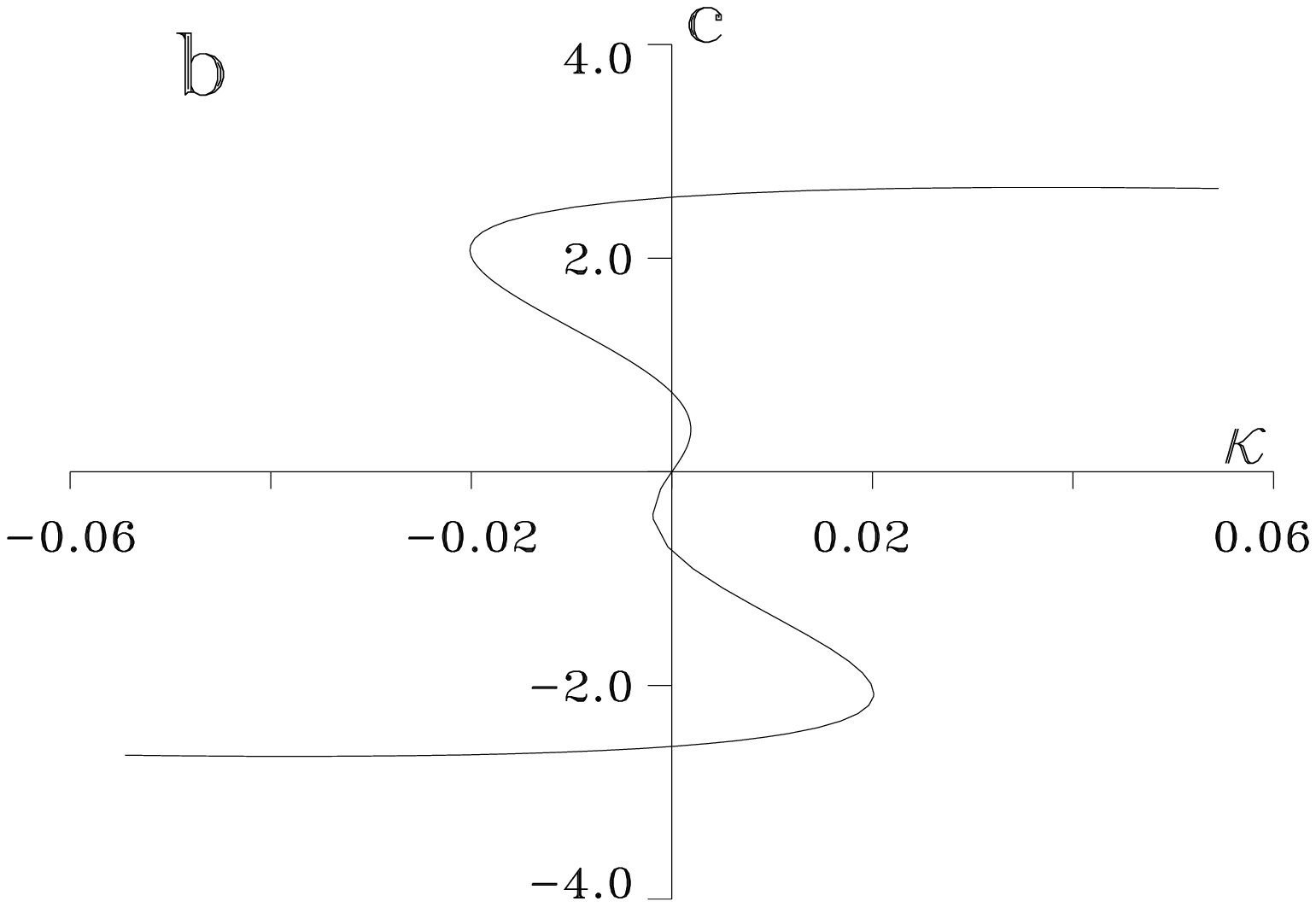}
}
\caption
{
Solutions to the speed vs curvature relation~(\protect\ref{cK}).
(a) Away from the bifurcation point (top), 
the speed $c$ varies approximately linearly
with the curvature $\kappa$ ($\epsilon=0.003$).
(b) Near the bifurcation point, small curvature variations may
drive the system past the endpoint of the solution branch and
cause pulse collapse ($\epsilon=0.022$).  Other parameters: $a_1=1.25$,
$a_0=-0.2$, $\delta=2.5$.
}
\label{fig:cvk}
\end{figure}

For the model of Eqns.~(\ref{pde1}),
the effect of curvature $\kappa$ on pulse propagation can be obtained
from Eqn.~(\ref{c}) by rewriting the equations
in a frame moving with the pulse~\cite{HaMe:97}. 
Assuming the radius of curvature is much 
larger than the pulse width, and a negligible dependence of $u$ and $v$
on arclength and time (in the moving frame) we obtain
\begin{eqnarray}
\label{ode}
\delta^{-1}u^{\prime\prime}+(c+\delta^{-1}\kappa)u^\prime+
\epsilon^{-1}(u-u^3-v)&=&0\,, \\
v^{\prime\prime}+(c+\kappa)v^\prime+u-a_1v-a_0&=&0 \,, \nonumber
\end{eqnarray}
where the prime denotes differentiation with respect to a coordinate 
normal to the front line.
Rescaling $\epsilon$ and 
$\delta$ by the factor $\frac{c+\delta^{-1}\kappa}{c+\kappa}$ 
Eqn.~(\ref{c}) becomes
\begin{equation}
\label{cK}
c+\kappa={\cal C}(\lambda;\frac{c+\delta^{-1}\kappa}{c+\kappa}
\eta,a_0,a_1)\,.
\end{equation}
Figure~\ref{fig:cvk} 
shows numerical solutions of Eqn.~(\ref{cK}) for $c$ in terms of
$\kappa$.  Far away from failure of propagation (Fig.~\ref{fig:cvk}a)
we find the usual approximate linear $c-\kappa$ relations for right ($c>0$) 
and left ($c<0$) propagating pulses~\cite{TyKe:88,Meron:92}. 
Close to failure (Fig.~\ref{fig:cvk}b), 
small realizable curvature variations may cause collapse.

\begin{figure}
{\center 
\includegraphics[width=3.25in]{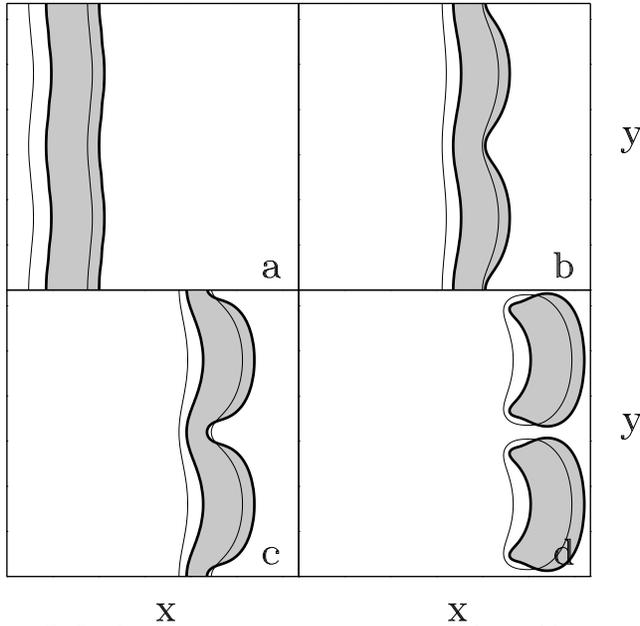}
}
\caption{Breakup of a pulse by transverse instability.  The thick and thin 
lines pertain to $u=0$ and $v=0$ contour lines, respectively. The initial
almost planar pulse is unstable to transverse perturbations and forms
a dent.  The dent grows and the pulse breaks at the region of
high curvature.
The equation parameters are the same as in Fig.~\protect\ref{fig:cvk}.
}
\label{fig:ksim}
\end{figure}

Equation (\ref{cK}) contains information also about the transverse stability 
of a pulse line. A positive slope of a $c-\kappa$ relation at $\kappa=0$ 
indicates an instability of a planar pulse. Fig.~\ref{fig:ksim} shows a 
simulation of Eqns.~(\ref{pde1}) at parameter values pertaining to the 
$c-\kappa$ relation in Fig.~\ref{fig:cvk}b.
Starting with an near planar pulse, dents grow due to a
transverse instability. The negative curvature that develops induces a wave 
breakup.

%
%
\section{Conclusion}

We have identified a mechanism for breakup of waves in an
excitable media. The key ingredient of this mechanism is the proximity to a
subcritical pitchfork pulse bifurcation 
(as shown in Fig.~\ref{fig:subcritical}a).
Near the bifurcation small perturbations become significant 
and may induce failure of propagation. 
The nature of the perturbation is of secondary importance. 
As illustrated in  Figs.~\ref{fig:cvj}b and 
\ref{fig:cvk}b, the effects of an
advective field and  curvature are similar; they both 
induce wave breakup by driving the system past the end points of propagating 
pulse branches.  A perturbation inducing breakup can be externally applied, 
like an electric
field in the BZ reaction, or spontaneously formed, like curvature growth by a
transverse instability.  An interesting question not resolved in this 
study is the observed preference of propagation failure or collapse 
rather than reversal in the direction of propagation. 

\acknowledgments
This study was supported in part by grant  No 95-00112 from the US-Israel
Binational Science Foundation (BSF).

\bibliography{reaction}

\end{document}